\documentclass[
twocolumn,
showpacs,showkeys,preprintnumbers,aps,prl]
{revtex4}

\usepackage{amsmath}
\usepackage{amssymb}
\usepackage{revsymb}
\usepackage{graphicx}
\usepackage{dcolumn}

\begin{document}

\title{Poincar\'e-invariant spectator-model currents
for electroweak nucleon form factors}
\author{R.F.~Wagenbrunn, T. Melde, W. Plessas}
\affiliation{Theoretische Physik, Institut f\"ur Physik, Karl-Franzens-Universit\"at \\
Universit\"atsplatz 5, A-8010 Graz, Austria}
\begin{abstract}
The relativistic constituent quark model of low-energy quantum chromodynamics
is found to yield a consistent picture of the electroweak structure of the
nucleons. Notably, the electromagnetic and axial form factors of both the
proton and the neutron can be described
in close agreement with existing experimental data in the domain of low
to moderate momentum transfers. For the theory it is mandatory to respect
Poincar\'e invariance and to fulfill additional conditions like charge
normalization. Here we present covariant predictions of the
one-gluon-exchange and Goldstone-boson-exchange constituent quark models
for the electroweak form factors of the nucleons and give a critical discussion
of the results in view of the point-form spectator model employed for the
electromagnetic and axial current operators.
\end{abstract}

\maketitle
The explanation of the electroweak structure of the nucleons
and of other baryon ground states still represents
a formidable problem. Even though the theoretical framework appears to be
well founded in the standard model of strong and electroweak interactions
and a wealth of experimental data has been accumulated up till now, one has
not yet reached a conclusive understanding of the electroweak and axial form
factors, especially of the nucleons, at low and intermediate energies. Of
course, the essential difficulties reside in the solution of quantum
chromodynamics (QCD) outside the perturbative regime.
Recently, calculations of the nucleon electroweak form factors have become
available from relativistic constituent quark models (RCQMs). The pertinent
covariant results have turned out remarkable in several respects.
The RCQM
assumes the nucleons to consist of three constituent quarks and describes
their mass spectra in a
Poincar\'e-invariant manner. The theory is formulated along relativistic
Hamiltonian dynamics. Thus one works a-priori with a finite number of degrees
of freedom rather than facing problems, such as regularization and truncation,
in a field-theoretical approach. The symmetries of Lorentz invariance are
strictly included by fulfilling the constraints of the Poincar\'e algebra.
Similarly the essential properties of (nonperturbative) QCD, like the
consequences of spontaneous breaking of chiral symmetry at low energies,
can be implemented in the Hamiltonian or equivalently in the invariant mass
operator defining the RCQM.

\begin{figure*}
\begin {center}
\includegraphics[width=7.5cm,clip=]{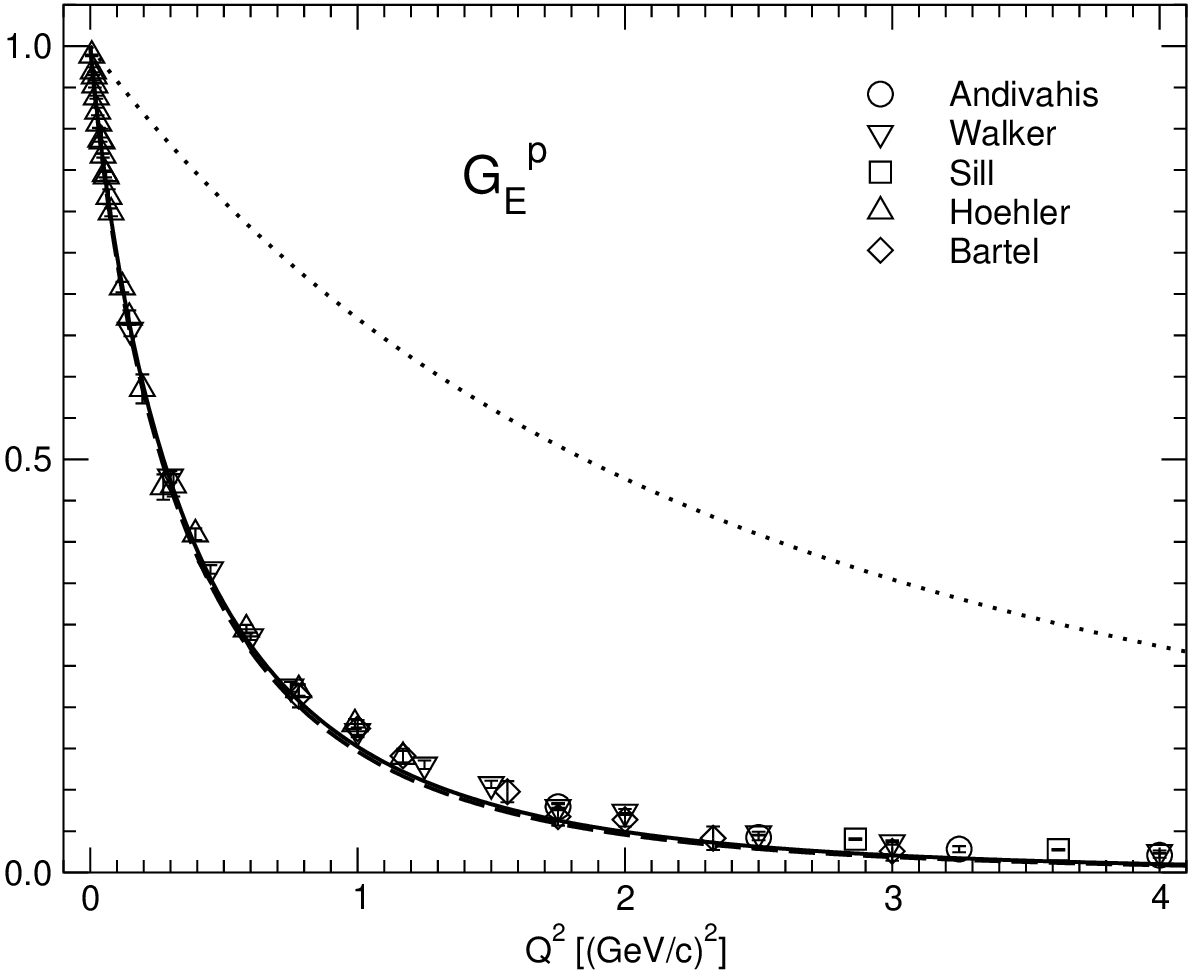}
\hspace{0.4cm}
\includegraphics[width=7.5cm,clip=]{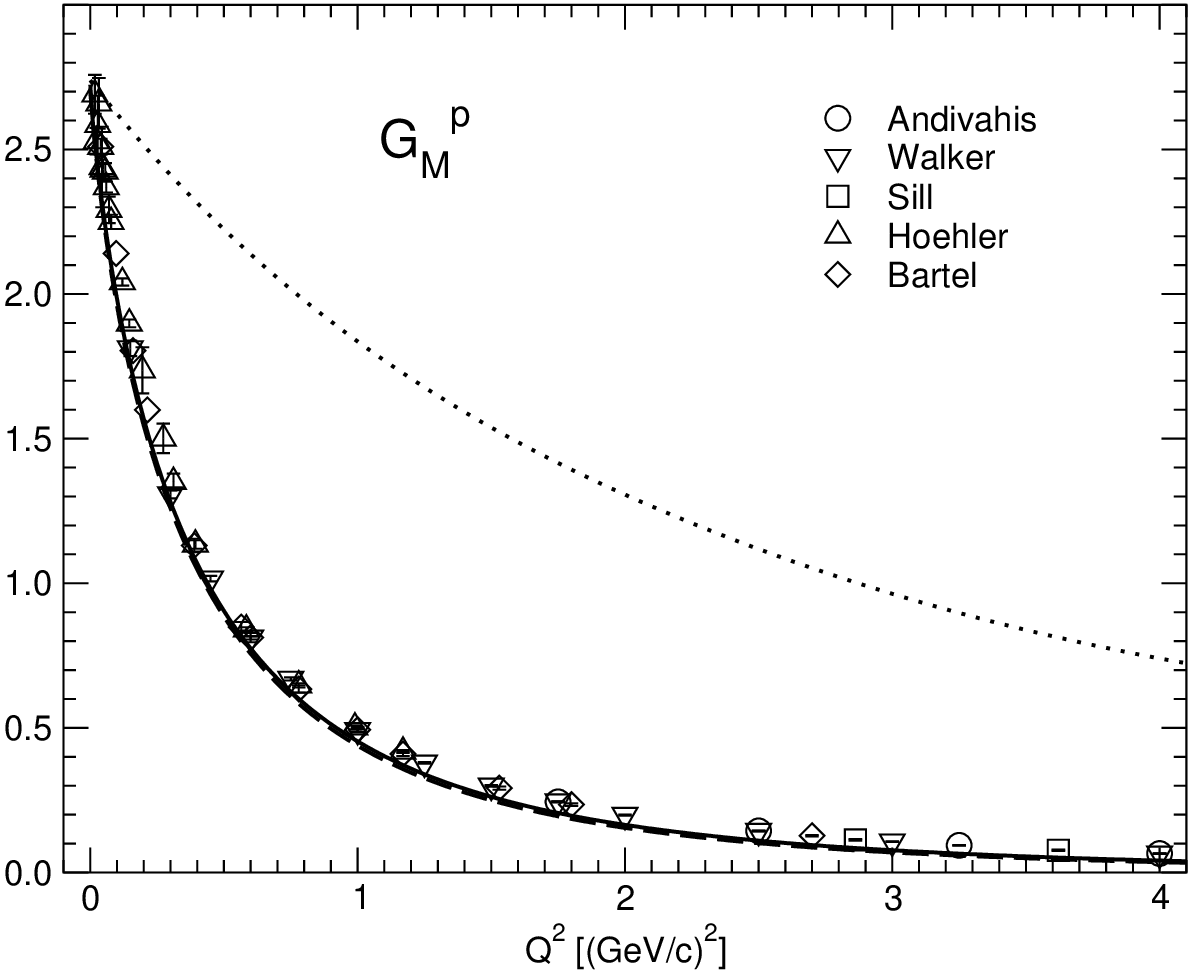}
\caption{\label{fig:proton}
PFSM predictions for the elastic electric and magnetic form factors of the proton
by the OGE and GBE relativistic constituent quark models (dashed and solid lines,
respectively). For comparison, the NRIA
results of the GBE CQM are also shown (dotted lines). Experimental data from
refs.~\cite{Andivahis:1994rq,Walker:1989af,Sill:1993qw,Hohler:1976ax,Bartel:1973rf}.
}
\end{center}
\end{figure*}

\begin{figure*}
\begin{center}
\includegraphics[width=7.5cm,clip=]{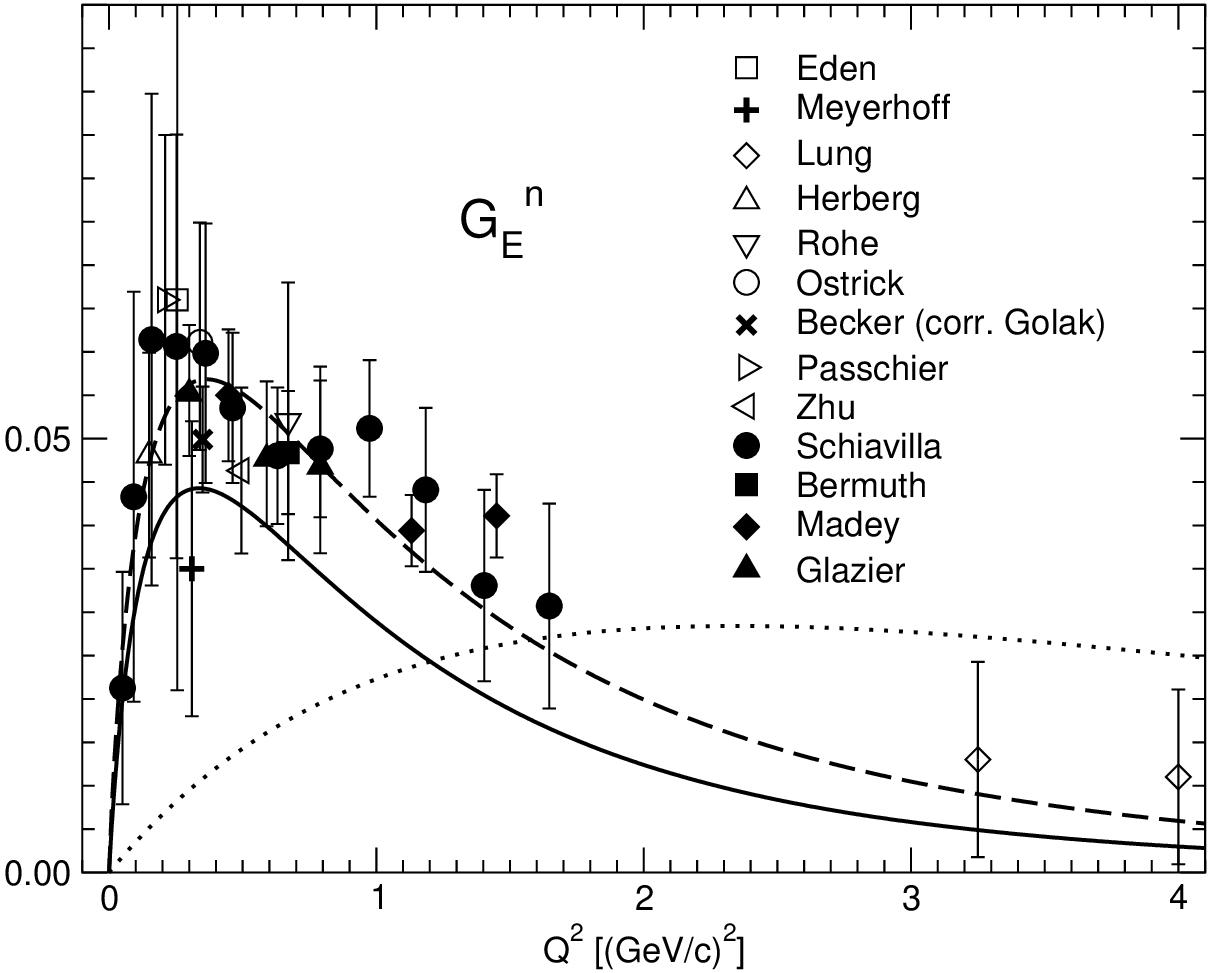}
\hspace{0.4cm}
\includegraphics[width=7.5cm,clip=]{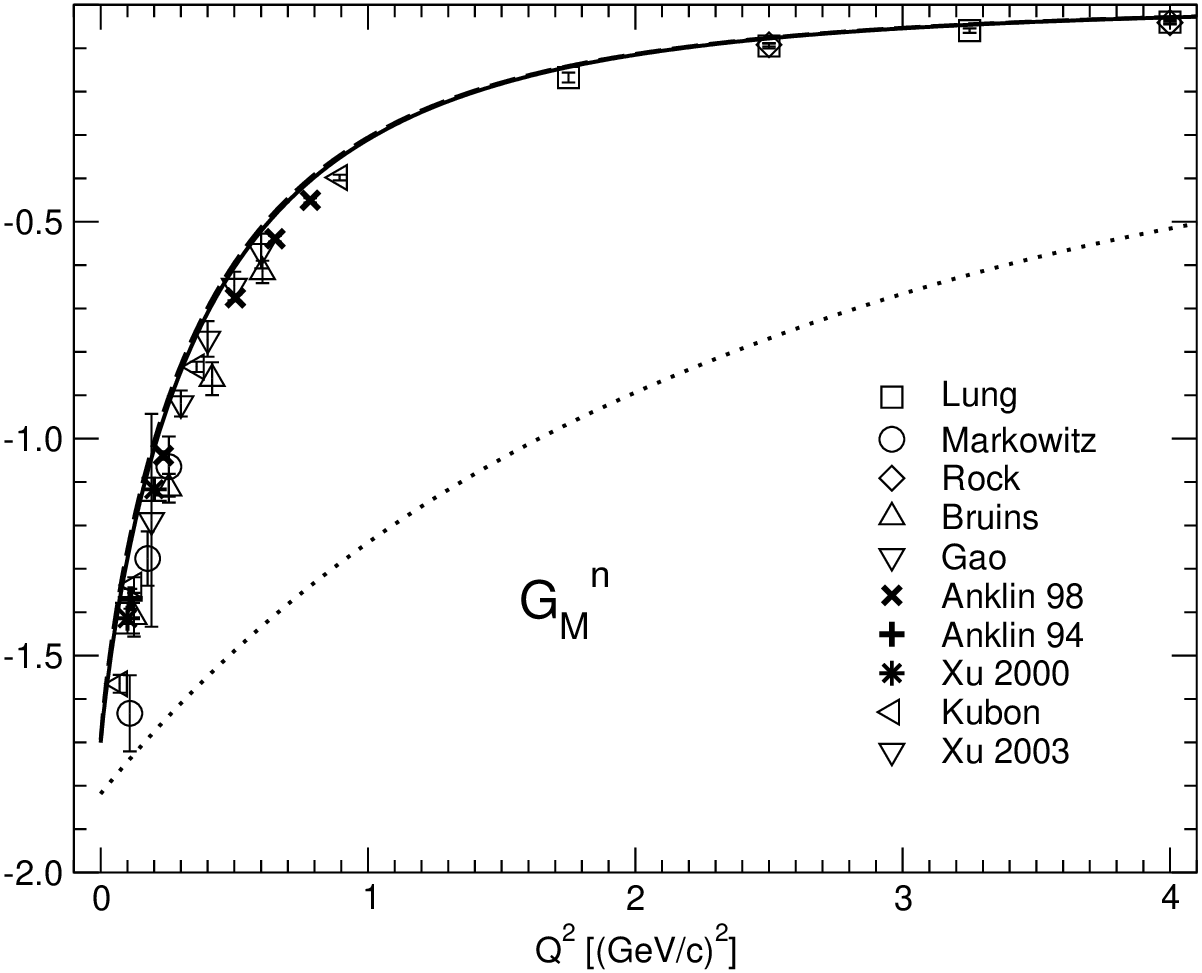}
\caption{\label{fig:neutron}
Same as Figure~\ref{fig:proton} for the neutron. Experimental data from 
refs.~\cite{Eden:1994ji,Meyerhoff:1994ev,Lung:1993bu,Herberg:1999ud,Rohe:1999sh,%
Ostrick:1999xa,Becker:1999tw,Passchier:1999cj,Zhu:2001md,%
Schiavilla:2001qe,Bermuth:2003qh,Madey:2003av,Glazier:2004ny,%
Markowitz,Rock,Bruins,Gao,Anklin,Anklin2,Xu,Kubon,Xu2}.
}
\end{center}
\end{figure*}

\begin{figure}
\begin{center}
\includegraphics[width=7.5cm,clip=]{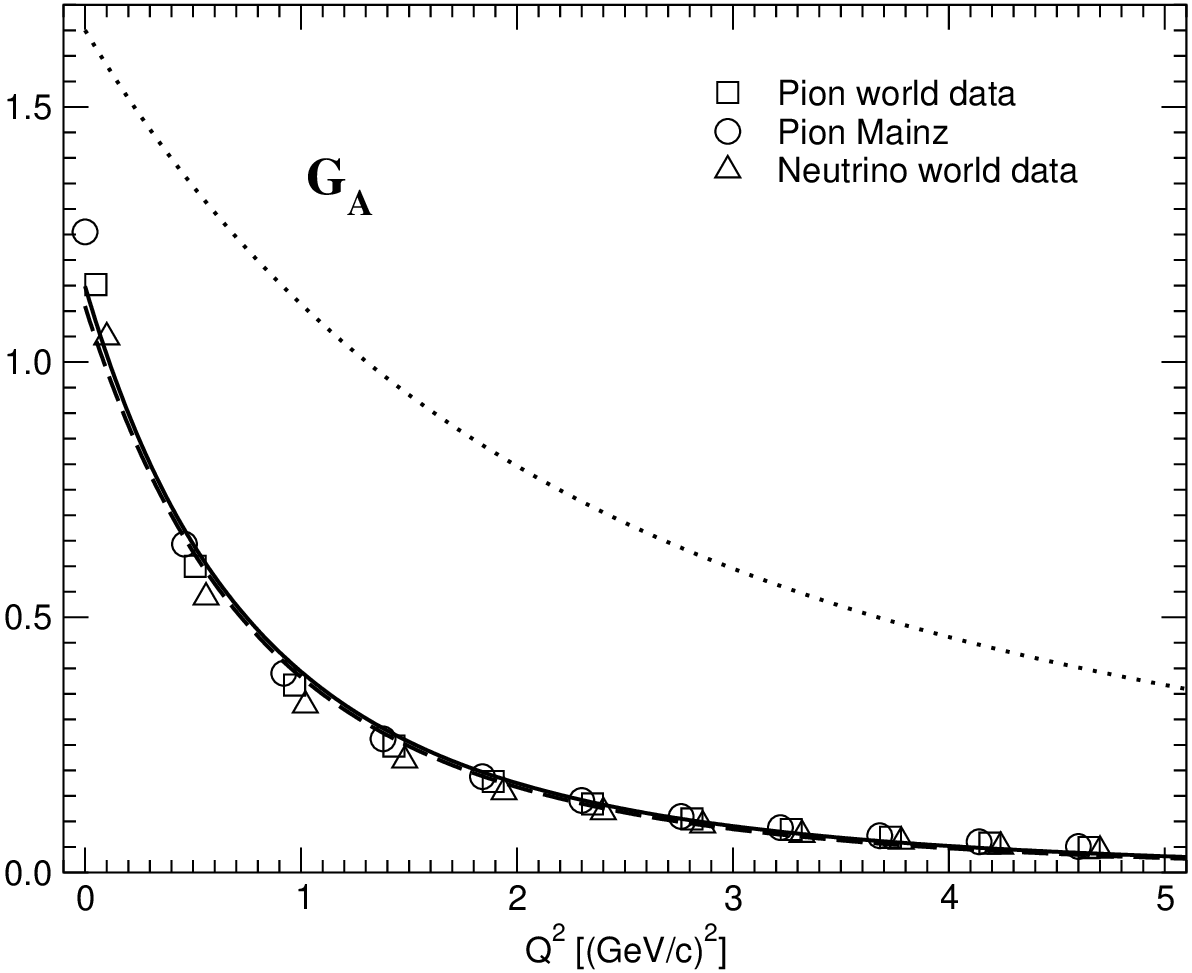}
\caption{\label{fig:axial}
PFSM predictions for the nucleon axial form factor. Same notation as in
Figure~\ref{fig:proton}. Experimental data are represented as dipole
parametrizations as in ref.~\cite{Glozman:2001zc}. 
}
\end{center}
\end{figure}

Elastic electroweak nucleon form factors have recently been studied in the
framework of relativistic quantum mechanics. Specifically, working within
the point form \cite{Dirac:49,Klink:98} allows to calculate all the desired
observables in a covariant manner. One has first produced the predictions of
the Goldstone-boson-exchange (GBE) RCQM \cite{Glozman:1997ag} for the
electroweak structure of the nucleons using spectator-model currents
\cite{Wagenbrunn:2000es,Glozman:2001zc,Boffi:2001zb,Berger:2004yi}. The
electromagnetic and axial nucleon form factors were readily found to be
remarkably consistent with all experimental data at low momentum transfers.
Here, we add the corresponding predictions of the one-gluon-exchange (OGE) RCQM
by means of the relativistic version of the
Bhaduri-Cohler-Nogami (BCN) model \cite{Bhaduri:1981pn} as parametrized in ref.
\cite{Theussl:2000sj}. The results are collected in 
Figures~\ref{fig:proton}-\ref{fig:axial}. There a comparison is given to the
predictions of the GBE RCQM, to the nonrelativistic impulse approximation (NRIA)
\cite{Wagenbrunn:2000es,Glozman:2001zc,Boffi:2001zb},
and to the existing experimental data. One observes only minor differences between
the RCQMs with different dynamics for the hyperfine interaction (the solid and
dashed curves in Figures~\ref{fig:proton}-\ref{fig:axial} are practically
indistinguishable, except for the
neutron electric form factor $G_E^n$). The reason is that
the relevant components of the nucleon ground-state wave functions are rather
similar for both the OGE and GBE RCQMs. The differences between the two types
of RCQMs become striking only for the excited states \cite{Glozman:1998fs}.
From the form factor
results in Figures \ref{fig:proton}-\ref{fig:axial} it is also immediately
evident that relativistic effects are of paramount importance in all respects.
The NRIA fails completely. 

In view of the existing results, one must ask why such a consistent picture
of the electroweak structure of the nucleons can really come about by employing
RCQMs; even more so since a simplified model
of the electromagnetic and axial currents has been employed. Up till now the full
many-body character of the electroweak currents cannot be tackled in a fully
relativistic calculation. Rather one has been resorting to simplifications.
The results exemplified here have been obtained within the so-called
point-form spectator model (PFSM) for the electromagnetic and axial currents.
The PFSM is characterized by the fact that the intermediate
boson couples only to one of the constituent quarks in the baryon, while the
momentum is transferred to the baryon as a whole. 
Among the available forms of relativistic quantum mechanics, the point form
is specific in the respect that the approximated current operators
preserve their spectator character in all reference frames \cite{MCPW:2005}.
One uses a covariant (spectator-model) current operator and in addition
further symmetry requirements, like translational and time-reversal
invariance as well as charge normalization, are implemented in the construction. 

\begin{figure*}
\begin{center}
\includegraphics[clip=,width=7.5cm]{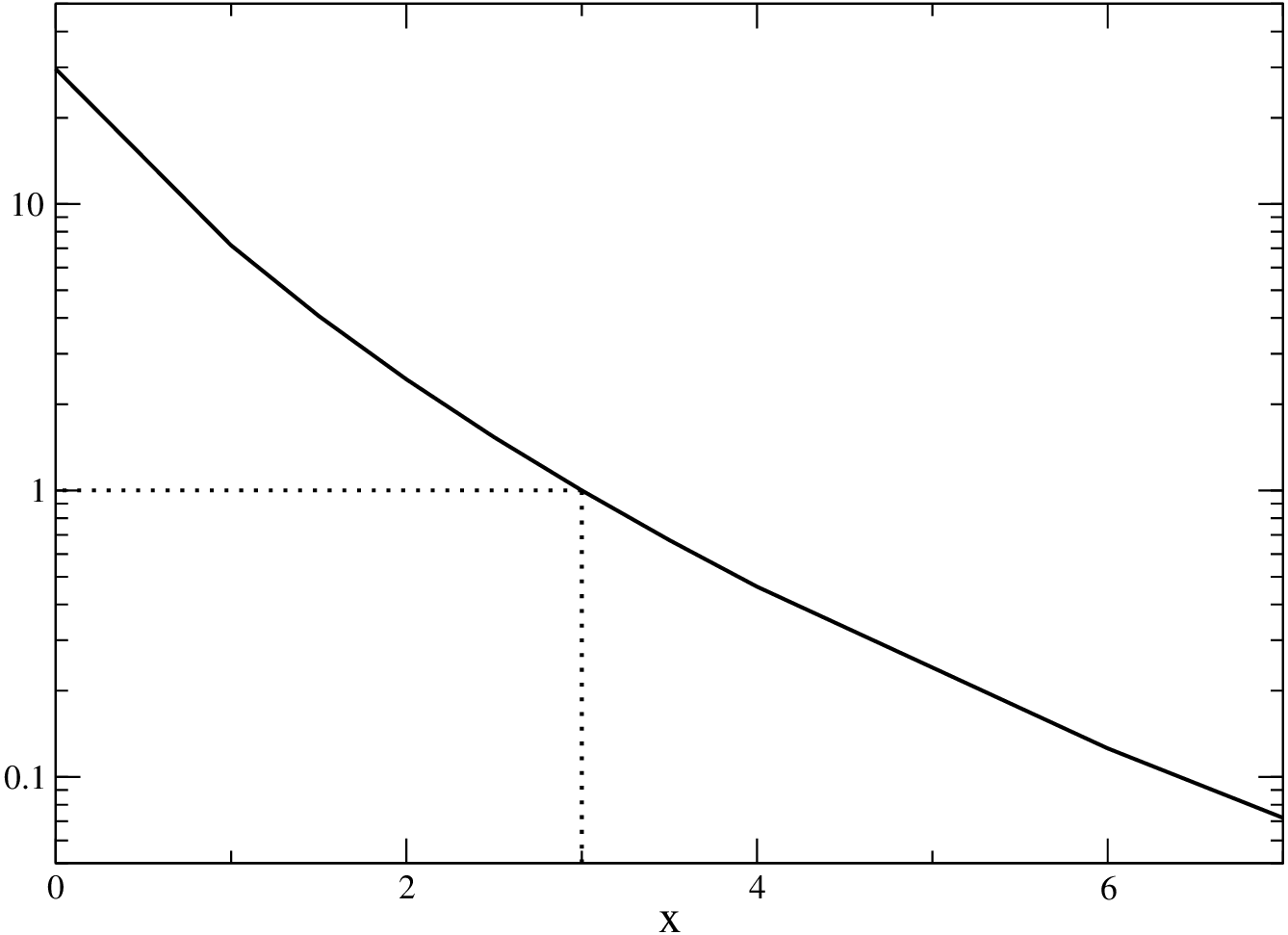}
\hspace{0.4cm}
\includegraphics[clip=,width=7.5cm]{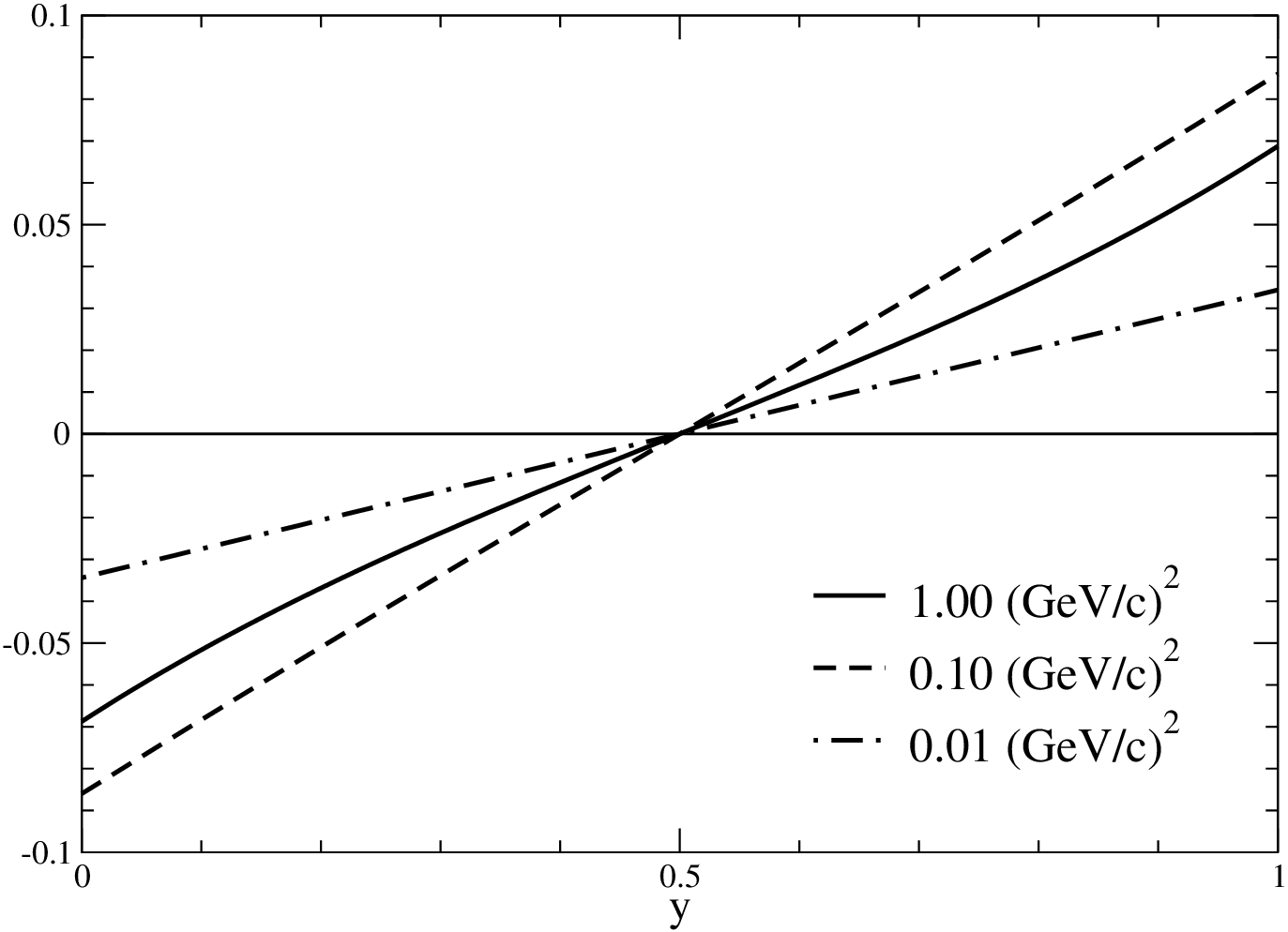}
\caption{\label{fig:chargenorm}
Left: Proton charge as a function of the exponent $x$ in the normalization factor
$ N$ of eq.~(\ref{eq:offsymfac}).
Right: Expectation value of the electromagnetic current component $\hat J^{\mu=3}$
in the Breit frame as a function of the exponent $y$ in the normalization factor
$ N$ of eq.~(\ref{eq:offsymfac}) for three different values of the momentum
transfer $Q^2$.
}
\end{center}
\end{figure*}

The concrete expressions for the reduced matrix elements of the PFSM current
between three-body states of quarks with individual
momenta $p_i$ and spin projections $\sigma_i$ read
\begin{multline}
\left<p'_1,p'_2,p'_3;\sigma'_1,\sigma'_2,\sigma'_3\right|
{\hat J}^{\mu}_{\rm rd}
\left|p_1,p_2,p_3;\sigma_1,\sigma_2,\sigma_3\right>
=\\
3 {N}
\left<p'_1,\sigma'_1\right|\hat{J}^{\mu}_{\rm spec}\left|p_1,\sigma_1\right>
\\
2p_{20}\delta^3\left({\vec p}_2-{\vec p}'_2\right)
2p_{30}\delta^3\left({\vec p}_3-{\vec p}'_3\right)
 \delta_{\sigma_{2}\sigma'_{2}}
   \delta_{\sigma_{3}\sigma'_{3}}
\label{eq:pfsm}\, .
\end{multline}
Here, the matrix element of the current operator between one-body states
of the constituent quark coupling to the intermediate boson is taken of
the form
\begin{equation}
\left<p'_1,\sigma'_1\right|\hat{J}^{\mu}_{\rm spec}\left|p_1,\sigma_1\right>=
e_1{\bar u}\left(p'_1,\sigma'_1\right)
\gamma^\mu
u\left(p_1,\sigma_1\right) \, ,
\label{eq:emcurr}
\end{equation}
for the electromagnetic case, and
\begin{multline}
\left<p'_1,\sigma'_1\right|\hat{J}^{\mu}_{a,{\rm spec}}
\left|p_1,\sigma_1\right>=
\\
{\bar u}\left(p'_1,\sigma'_1\right)
\left[g_A^q \gamma^\mu 
+ \frac{2f_\pi}{\widetilde{Q}^2+m_\pi^2} g_{qq\pi}\widetilde{q}^\mu\right]
\gamma_5 \frac{1}{2}{\tau}_a u\left(p_1,\sigma_1\right) \, ,
\label{eq:axcurr}
\end{multline}
for the axial case, where $u\left(p_1,\sigma_1\right)$ represents the four-component
Dirac spinor of quark 1; for more details of the formalism see ref.~\cite{MCPW:2005}.
Obviously, eq.~(\ref{eq:emcurr}) represents the usual relativistic electromagnetic
current for a point-like particle, and eq.~(\ref{eq:axcurr}) is the conventional
axial current with the pion-pole term included. Now, there are several important
features to be noted about the PFSM currents.
First, all four components of the momentum transfer
$
{\tilde q}^{\mu}={p}^{\mu}_1-{p'}^{\mu}_1
$
to the struck quark are in general different from the one of the momentum 
transfer $q^\mu$ to the baryon as a whole; $\tilde q^{\mu}$ results uniquely
from the spectator conditions in eq.~(\ref{eq:pfsm}) and from the
conservation of the overall momentum ${q}^{\mu}={P}^{\mu}-{P'}^{\mu}$
(translational invariance). 
Second, in the PFSM construction of the current in 
eq.~(\ref{eq:pfsm}) a normalization factor ${ N}$ has to be introduced in order
to guarantee for the proper charge normalization (of the
proton)\footnote{It should be emphasized that this normalization factor in the current
operator has nothing to do with the normalization of the nucleon states. The
necessity of employing $N$ in the PFSM current operator could easily be overlooked.
It becomes especially evident, however,
when correctly normalized states are used in the evaluation of the matrix element
(\ref{eq:pfsm}), see the details given in ref.~\cite{MCPW:2005}.}. 
In principle, a Lorentz-invariant form of the normalization factor ${ N}$
involves the interacting masses $M$ and $M'$ of the incoming and outgoing baryon
states, respectively, and can be chosen in many different ways \cite{MCPW:2005}
\begin{equation}
{ N}\left(x,y\right)=
\left(\frac{M}{\sum_i{\omega_{i}}}\right)^{xy}
\left(\frac{M'}{\sum_i{\omega'_{i}}}\right)^{x\left(1-y\right)}\, ,
\label{eq:offsymfac}
\end{equation}
with $0 \le x$ and $0\le y \le 1$. 
From the electromagnetic case, the exponent $x$
is fixed to 3 so that the electric form factor $G_{E}^p$ yields the proper
charge of the proton, see the left panel of Figure~\ref{fig:chargenorm}.
The exponent $y$ can be constrained specifically by exploiting time-reversal
invariance. The latter
implies that in the Breit frame the expectation value of the third component
$\hat J^{\mu=3}$ of the current operator has to vanish~\cite{Durand}. From the
behaviour of $N$ as a function of $y$ in the right panel of
Figure~\ref{fig:chargenorm} one finds that $y=\frac{1}{2}$ meets this constraint
(for all values of the momentum transfer $Q^2$).
It should be noted that the normalization factor $N$ entering into
eq. (\ref{eq:pfsm}) introduces contributions from the interacting three-quark
systems in a non-separable manner and thus makes the PFSM currents effective
many-body ones. 

In view of the constraints implemented in its construction, the PFSM is found
to be specific, since it does not only guarantee for the invariance of the
transition amplitudes under the transformations of the whole Poincar\'e group
(including space and time reflections as well as space-time translations) 
but also allows to fulfill
supplementary requirements such as charge normalization. All of these
constraints are maintained in any reference frame,
because the point-form calculations are performed in a manifestly covariant manner. 
One may suspect that the relatively good performance of the PFSM approach is
due to the fulfilling of these additional conditions beyond Poincar\'e invariance.
\begin{acknowledgments}
This work was supported by the Austrian Science Fund (Project P16945-N08).
\end{acknowledgments}
%

\end{document}